\documentclass[%
 reprint,
bibnotes,
 amsmath,amssymb,
 aps,
prd,
intlimits,sumlimits,
]{revtex4-1}

\usepackage{natbib}
\usepackage{tikz}
\usetikzlibrary{matrix}
\usepackage{enumerate}

\def\HH{\mathsf{H}} 
\def\HHG{\mathsf{H_G}} 
\def\HHF{\mathsf{H_F}} 
\def\PP{\mathsf{P}} 
\def\PPF{\mathsf{P}_{\hspace{-0.03cm}\mathrm{F}}} 
\def\PPG{\mathsf{P}_{\!\mathrm{G}}} 

\def\MM{\mathsf{M}}

\def\BH{\mathrm{BH}} 

\DeclareMathOperator{\Tr}{\mathrm{Tr}}

\DeclareMathOperator{\met}{\mathrm{m}}

\def\pd{\partial} 
\def\eps{\varepsilon} 
\newcommand{\ket}[1]{| #1 \rangle} 
\newcommand{\bra}[1]{\langle #1 |} 

\newcommand{\avg}[1]{\langle #1 \rangle}

\DeclareMathOperator{\Span}{\mathrm{span}} 
\DeclareMathOperator{\ii}{\mathrm{i}}

\def\doublesumone{\sum_{a=0}^{N_G-1} \sum_{n=0}^{q_1-1}}
\def\doublesumtwo{\sum_{a=0}^{N_G-1} \sum_{n=0}^{q_2-1}}

\begin{document}

\title{On the implications of the Bekenstein bound for black hole evaporation}

\author{Giovanni Acquaviva}
\email{gioacqua@utf.troja.mff.cuni.cz}

\author{Alfredo Iorio}
\email{iorio@ipnp.troja.mff.cuni.cz}

\author{Martin Scholtz}
\email{scholtz@utf.mff.cuni.cz}

\affiliation{Faculty of Mathematics and Physics, Charles University in Prague, V Hole\v{s}ovi\v{c}k\'ach 2,
18000 Prague, Czech Republic}

\begin{abstract}
We provide general arguments regarding the connection between low-energy theories (gravity and quantum field theory) and a hypothetical fundamental theory of quantum gravity, under the assumptions of \emph{(i)} validity of the holographic bound and \emph{(ii)} preservation of unitary evolution at the level of the fundamental theory.  In particular, the appeal to the holographic bound imposed on generic physical systems by the Bekenstein-Hawking entropy implies that both classical geometry and quantum fields propagating on it should be regarded as phenomena emergent from the dynamics of the fundamental theory.  The reshuffling of the fundamental degrees of freedom during the unitary evolution then leads to an entanglement between geometry and quantum fields.  The consequences of such scenario are considered in the context of black hole evaporation and the related information-loss issue: we provide a simplistic toy model in which an average loss of information is obtained as a consequence of the geometry-field entanglement.
\end{abstract}


\maketitle

\section{Introduction}
\label{sec:introquasiparticles}

In this paper we shall combine two general considerations regarding quantum theories of gravity and study their implications for black hole evaporation.

The first consideration is that, if the Bekenstein upper bound on the entropy of any physical system is correct, there probably
exist more fundamental entities than the ones we deem to be elementary \cite{tHooft1993,Susskind1995,Bousso2002}. The second consideration is that, at our energy scales,
these fundamental entities must organize themselves as quantum fields acting on classical spacetimes
(our best understanding of the low-energy physics) by ``making'' both the quantum fields and the classical geometry.

The consequence of such assumptions is then the following: since the bound is reached only when the black hole is formed, and it is
an \emph{upper} bound, then and only then \emph{all} the degrees of freedom of the fundamental entities have been excited.
Hence one can think of the black hole state as a gas-like high-energy ``phase'', with few (one in the simplest case) macroscopic parameters characterizing all
the microscopic states. When the energy is lowered (that is after the evaporation of the black hole) these fundamental entities have at their disposal a large amount of different, nonequivalent rearrangements. These are their physically nonequivalent low-energy ``phases'', that 
can only be described as specific quantum fields acting on specific geometries. Similar lessons can be learned from ordinary states of matter, made of ordinary particles. On the other hand there is a crucial difference with the usual phases of matter, namely the fact that we introduce here a democracy between spacetime and fields/particles which, in this view, both emerge from one underlying dynamics. We call this view the ``quasiparticle'' picture, as we shall explain below.

With this in mind, it is clearly virtually impossible that after the black hole evaporation we can retrieve the very same ``phase''  we had before the black hole was formed. Hence, the information associated to the quantum fields in the ``phase'' before the formation of the black hole is, in general, only partially recovered in the
``phase'' after the black hole has evaporated; the information loss is due to the entanglement between the fields and the geometry. This is how we intend to address the information paradox. 

The structure of the paper is as follows. In Sec.\ \ref{sec:quasi} we explain more in detail the quasiparticle picture and its role in the emergence of
both quantum fields and classical geometry. In Sec.\ \ref{sec:page-curve} we briefly present the construction of the Page curve as well as analogous concepts
and antagonist views. In order to demonstrate possible implications of our picture, in Sec.\ \ref{sec:model} we shall construct a toy model of
black hole evaporation that exhibits partial loss of information and, hence, leads to a modification of the Page curve.

\section{Quasiparticle picture}
\label{sec:quasi}

As widely known, the entropy $S$ of any physical system contained in a volume $V$, including the volume itself, is supposed to be bounded from above by the value of
the Bekenstein-Hawking entropy associated to a black hole whose event horizon coincides with the boundary of $V$~\cite{Bousso1999}
\begin{align}\label{eq:BekensteinBound}
	S \leq S_{\BH} = \frac{1}{4} \frac{\pd V}{\ell^2_P},
\end{align}
where $\ell_{P} = \sqrt{\hbar\, G / c^3} \sim 1.6 \times 10^{-35} \met$. The generality of the original bounds for ordinary matter (i.e.,\
when gravity is not included) posited by \cite{Bekenstein1981} is the subject of intense investigation and
debates \cite{Bekenstein2014}. Nonetheless, it is widely accepted that for black holes the upper bound is saturated.

By the number of degrees of freedom $N$ of a quantum physical system we mean \emph{the number of bits of information necessary
to describe the generic state of the system}. In other words, $N$ is the logarithm of $\cal{N}$, the dimension of the
Hilbert space of the quantum system. In the extreme case of a black hole ${\cal N} = e^{S_{\BH}}$. Hence, formula \eqref{eq:BekensteinBound} means \emph{i)} that in
nature the information contained in any volume $V$ cannot exceed 1 bit every 4 Planck areas of the boundary of $V$, and \emph{ii)} that
 \emph{if and only if} a black hole is formed, the degrees of freedom of the hypothetical fundamental entities are most excited (see, e.g., \cite{Bekenstein2003}).

These fundamental entities cannot fully coincide with the particles customarily
thought of as elementary (electrons, neutrinos, photons, etc.) for two reasons: one is that, if it were so, we would reach
the bound with ordinary matter, and this does not happen; the second one is that gravity must be included in the counting of the fundamental degrees of freedom because the
saturation only happens when gravity becomes as important as the other interactions.

Although we do not know the dynamics generating the fundamental degrees of freedom,
such dynamics needs be such that the emergent behavior at typical wavelengths much bigger than the Planck length $\ell_{P}$ is
that of continuum quantum fields acting on a continuum classical spacetime. That is, at low resolution we have quantum
field theory (QFT) in curved spacetime, hence both (classical) geometry and (quantum) fields are emergent entities.

On the one hand, the idea that gravity is an emergent phenomenon arising from more fundamental degrees of freedom is not new and goes back to
Shakarov \cite{Sakharov1967,Visser2002}. Presently there exist many particular models describing how gravity could emerge. The common
feature of these models is to consider some kind of underlying discrete lattice. A striking
fact that crystals with deffects can give rise to effective non-Euclidean geometry has been employed in the cosmological ``world crystal model'' \cite{Kleinert1987}.
It was proposed in \cite{VanRaamsdonk2010} that the classical properties of the space-time might emerge from the quantum entanglement between the actual fundamental degrees
of freedom. A specific model along these lines has been proposed recently in \cite{Cao2016}. An interesting feature of this model is the
possibility to recover the ER=EPR conjecture \cite{maldacena2013,jensen2013}. In quantum graphity \cite{Konopka2006,Konopka2008},
fundamental degrees of freedom and their interactions are represented by a complete graph with dynamical structure. For more approaches,
see, e.g., \cite{Baez1999,Ambjorn2006,Lombard2016,Oriti2014,Rastgoo2016,Requardt2015,Trugenberger2016}.

On the other hand, emergent, nonequivalent descriptions of the same underlying dynamics are a built-in characteristic of QFT \cite{Haag1992}, 
both in its relativistic regime \cite{Dirac1966} (hence deemed to be fundamental) 
and in its nonrelativistic regime \cite{Umezawa1982} (e.g., in condensed matter). 
Indeed, it is well recognized by now that the quantum vacuum can have a rich structure \cite{Milloni2013}
with nonequivalent quantum mechanical sectors or ``phases''. This complexity is understood in QFT as due to the infinite number of degrees of freedom and/or to the nontrivial
topology of the system, such as the presence of topological defects \cite{Blasone2011}.

On the mathematical level these features are the manifestation of the failure of the Stone-von Neumann theorem \cite{Neumann1931,Hall2013} that holds only for quantum mechanical systems of
finite degrees of freedom and trivial topology \cite{Bogolubov2012}. This failure leads to the existence of different, unitarily inequivalent
representations of the field algebra. That is, for a given dynamics one should expect several different Hilbert spaces, representing
different ``phases'' of the system with distinct physical properties, and distinct excitations playing the role of
the elementary excitations \footnote{In fact, the concepts of elementary and collective excitations are interchangeable in theories where electromagnetic duality is at play \cite{Montonen1977,Castellani2016}.}
for the given ``phase'' \cite{Umezawa1993}, but whose general character is that of the quasiparticles
of condensed matter \cite{Landau8,Landau9}.

In condensed matter examples are many. From the Cooper pairs of type II superconductors \cite{BSC1957,Altland2010} that are bosonic quasiparticles emerging
from the basic fermionic dynamics of the electrons interacting with the lattice, to the more recently discovered quasiparticles of graphene 
\cite{CastroNeto2009} that are massless Dirac quasiparticles emerging from the 
dynamics of electrons propagating on carbon honeycomb lattices, and giving raise to a continuum relativistic-like (2+1)-dimensional field theory on a pseudo-Riemannian geometry.

Similarly, as well known, one finds examples also in the context of black hole physics, the one of interest here. Indeed, the vacuum of a freely falling observer in Schwarzschild's spacetime
can be seen, by a static observer, as a coherent state of Cooper-like pairs, similar to that of a superconductor \cite{Israel1992}, and 
the Hawking radiation itself is related to the existence of distinct elementary excitations in the two frames. See the original derivation 
of Hawking \cite{Hawking1974,Hawking1976}, and also \cite{Israel1976,Iorio2004}.

Those degrees of freedom, though, are not the fundamental ones we are referring to here, because they do not explicitly include the degrees of freedom of geometry
\footnote{In fact, in the case of graphene, geometries can indeed be seen as emergent \cite{Iorio2011,Iorio2015}. Actually, inspired by that fact that
  different arrangements of the carbon atoms can give rise to the same emergent spacetime geometry, in our model we take into account 
the possibility that the same emergent geometry can be realized through different arrangements of the fundamental degrees of freedom. These microscopic
arrangements are indistinguishable at our (low) energy level.}.  Such extra request is suggested by the Bekenstein bound, but the general mechanism we propose is similar to the one at work already in ordinary matter.

The bound \eqref{eq:BekensteinBound} does not identify the type of fundamental degrees of freedom nor their dynamics. Nonetheless, we can extract from that bound
one important consequence for the process of black hole evaporation. In the standard scenario assuming unitary evolution, the information contained in the collapsing
matter is scrambled inside the black hole, but is eventually fully released during evaporation. This paradigm of information conservation
is manifested by the so-called ``Page curve'' \cite{Page1993b} which describes the complete information retrieval in the Hawking radiation at the final stage
of the black hole evaporation. In our picture, however, the probability that after the complete evaporation the fundamental degrees of freedom reorganize just like before the collapse
leading to black hole, is inversely proportional to the number of possible nonequivalent
rearrangements of the fundamental degrees of freedom. Therefore, even if one demands the dynamics of the fundamental degrees of freedom to be unitary, as we shall do,
one expects that the entanglement between the geometry and the quantum fields due to the reshuffling of fundamental degrees of freedom could lead to an effective loss of information
in the Hawking radiation.

The loss of information, in the sense of evolution of a pure state into a mixed state, can have two causes. The first one is that the laws of quantum theory are indeed
violated in some regimes. The second one is that only some subsystem of the universe is accessible, hence there will always be a
residual entanglement of the subsystem with the inaccessible parts \cite{Wald-Unruh2017}. In our picture we do not consider the first possibility, rather we suggest
that part of the total system is always hidden: this produces entanglement between emergent fields and geometry and leads to an effective loss of information on the field side.

Let us conclude this introduction with a schematic summary of the logic behind the model we propose.

\begin{enumerate}[i)]

\item We interpret the upper bound \eqref{eq:BekensteinBound} as indication of the existence of finite number of fundamental degrees of freedom,
  fully excited (saturated bound) only for a black hole. To access these fundamental degrees of freedom, one would need resolutions of order $\ell_P$ (which might not be possible at all, as suggested, e.g.,\ in \cite{Doplicher1995}).

\item Everything we see at our low-energy scale (low-resolution) is classical spacetimes and quantum fields. Both emerge from the properties of and the interactions between fundamental degrees of freedom.

\item Since the bound is not reached at our energies, the particles we call elementary are in fact emergent quasiparticles.

\item Being discrete entities, the fundamental degrees of freedom must arrange into discrete structures. Their nature (symmetries, type of interaction, etc.) is not specified here, as 
we shall make model-independent considerations. In other words, we do not construct here a specific model of quantum gravity, but try to convey general considerations.

\item There are, in general, different configurations of the fundamental degrees of freedom which give rise to the same classical geometry. These configurations yield different
  numbers of degrees of freedom for the fields. Thus, even if the geometries before the formation of the black hole and after its evaporation are the same, the emerging
  quantum fields will be, in general, different (i.e., live in different Hilbert spaces).

\item Even though, for simplicity, we assume unitary evolution on the fundamental level, the rearrangement of the fundamental degrees of freedom during the evaporation process 
leads to an entanglement between the emerging geometry and the emerging fields, thereby producing a loss of information on the field side.
\end{enumerate}

\section{Page curve}\label{sec:page-curve}

Our primary motivation is to address the black hole information paradox, 
i.e.,\ the  problem of the apparent loss of information during the process of a black hole evaporation. 
There are many proposals how to resolve this paradox. There are arguments that in the presence of gravity,
and especially in the presence of a black hole, we have to expect some modifications of the quantum theory and,
perhaps, deviations from unitary evolution at the fundamental level. From this perspective, there is no paradox in losing information
during the formation of a singularity and the subsequent evaporation of the back hole
\cite{Penrose1996,Modak2015}, because the underlying theory does not require the information conservation.

On the other hand, it has been advocated by \cite{Thooft1990} and \cite{Susskind1993} that the evolution is always unitary and the information loss is prohibited.
These arguments rely on the holographic principle, string theory models of black hole evaporation and the paradigm of the black hole complementarity.
Another confirmation of information conservation has been provided by \cite{Hawking2004}, who employed the quantum perturbations of
the event horizon and the AdS/CFT correspondence in order to argue that information can, in fact, escape from the black hole. 

There have been also arguments that it is impossible to reconcile the unitary evolution, the principle of equivalence and the low energy effective 
quantum field theory.  These arguments have been embodied in the controversial ``firewall paradox'' introduced in \cite{Almheiri2013}. 
Very recently, it was proposed in \cite{Hooft2016} how to avoid the firewall paradox by appropriate identification of the antipodal points
of the event horizon. 

A more conservative approach to the problem has been adopted by Page and it is based on purely quantum mechanical considerations.
Following \cite{Page1993a}, one considers the splitting of a Hilbert space $\HH$ into a bipartite system, $\HH = \HH_A^m \otimes \HH_B^n$,
where superscripts $m$ and $n$ indicate the dimension of corresponding Hilbert space, so that $\dim \HH = m\,n$.
Next, one chooses an arbitrary fixed state $\ket{\psi_0} \in \HH$ and a random unitary matrix $U$;
then $U \ket{\psi_0}$ is a random state in $\HH$. To such state we associate the density matrix $\rho_A(U)$, by tracing
out the subsystem $B$, and the corresponding entanglement entropy $S_{m,n}(U)$.
Averaging through $U$ we get the average entanglement entropy of the subsystem $A$,
\begin{align}
  S_{m,n} &= \left\langle S_{m,n}(U) \right\rangle_{\text{average through $U$}},
\end{align}
 and the average information contained in $A$,
\begin{align}
  I_{m,n} &= \ln m - S_{m,n}.
\end{align}
For mathematical details of this construction see the original paper \cite{Page1993a}, and also \cite{Harlow2016}.
Page conjectured -- and it was later proved in \cite{Foong1994} -- that the average information is
\begin{align}\label{eq:page information}
  I_{m,n} &= \ln m + \frac{m-1}{2n} - \sum_{k=n+1}^{mn} \frac{1}{k}, \qquad \text{for}~m < n.
\end{align}

These results are applied to the black hole evaporation problem in \cite{Page1993b}.
It is assumed that the evolution of the collapsing matter to produce a black hole and the subsequent evaporation of that black
hole is a unitary process, and hence there exists a $S$-matrix relating the initial collapsing matter to the final state
when black hole is fully evaporated and only the Hawking radiation remains.
The Hilbert space of the Hawking radiation is factorized into a product as before, where the subsystem $A$ now corresponds
to the states under the horizon and the subsystem $B$ corresponds to photons already emitted from the black hole.

\begin{figure}
  \centering
  \includegraphics[width=0.5\textwidth]{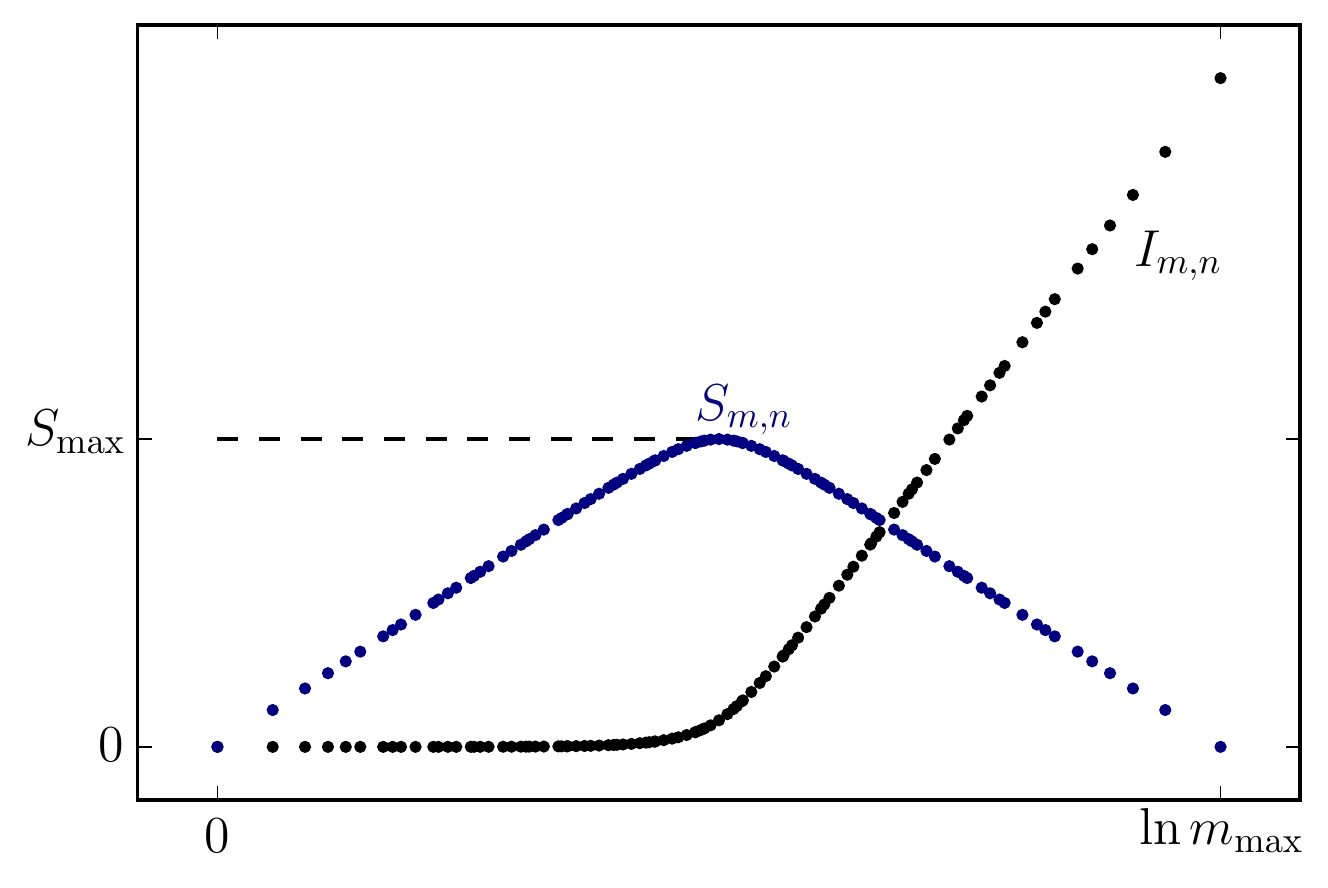} 
  \caption{Page's curve. The entanglement entropy between the subsystems and
  the information contained in the Hawking radiation are plotted against the
  dimension of the Hilbert space representing the emitted
  radiation. (Adapted from \cite{Page1993b}.)}
  \label{fig:page-original}
\end{figure}

When the black hole is formed, there is no Hawking radiation outside and, hence, $n = 1$ and $m = \dim \HH$.
Thus, by assumption, the entanglement entropy is trivially zero. As the black hole evaporates,
dimension $n$ increases and $m$ decreases, while $m \,n$ is kept constant. 
Since the emitted photons are entangled with the particles under the horizon, entanglement entropy increases.
At some stage of the evaporation (approximately half time of evaporation process) the information stored
below the horizon starts to leak from the black hole, decreasing the entanglement entropy. Finally, when the black hole
fully evaporates, $m = 1$ and $n = \dim \HH$ and the entanglement entropy returns to zero. 
The process is shown in Fig.\ \ref{fig:page-original}.

A natural generalization of the Page analysis is to consider tripartite system instead of a bipartite one.
In \cite{Hwang2016} the authors investigate the possibility that the particles emitted by
a black hole are transformed either into the Hawking radiation or into another form of matter,
which can be, e.g., a remnant. In this case, even when the black hole is fully evaporated, there can still
exist entanglement between these two forms of matter. Hence, the Hawking radiation does not contain the full information and it is not in a pure state (at least on average).

In the aforementioned works, no analysis of the interaction between the matter fields and the space-time geometry has been given, nor even addressed.
However, in order to consider the full black hole evaporation process, one certainly cannot treat matter as the test field on a given, say Schwarzschild, background. 
It is the system gravity+matter which evolves unitarily, not just the matter field.
Therefore, we propose the possibility that at the end of the evaporation the Hawking radiation is not in a pure state
because of its entanglement with the geometry itself and it is a purpose of this paper to clarify this statement.

Thus, in a sense, we proceed analogously to Page, and it is important to stress the points where we differ.
First, we interpret both gravity and fields as emergent phenomena. 
Thus, similarly to Page, we consider what we call \emph{fundamental} Hilbert space $\HH$,
but we do not split it into a direct product of the two spaces, 
because we claim that on the fundamental level there is no distinction between the field and the geometry at all.
Instead, we introduce \emph{effective} Hilbert spaces representing the states of the geometry and of the fields,
and mappings which extract the geometrical/matter content from the states of $\HH$.
Second, as a consequence, for the effective loss of the information we do not require the presence of a third,
unknown kind of matter like in \cite{Hwang2016}, because it is the entanglement of the field with the geometry which implies this effective loss.
Nonetheless, our description allows an arbitrary number of different fields, hence includes the possibility for such unknown kind of matter.

On the other hand, our approach is similar to that of Page in that we do not specify any particular microscopic dynamics.
We merely provide a kinematical framework which allows us to estimate the entanglement entropy.

\section{Model of black hole evaporation}\label{sec:model}

Our goal in this section is to construct a simple kinematical model which mimics the evaporation of the black hole, keeping in mind the illustrated conceptual framework
for which both geometry and quantum fields are emergent phenomena. We consider the following idealized scenario:
\begin{enumerate}
\item Initially, there is a quantum field (in an almost flat space) which collapses and eventually forms a black hole of mass $M_0$.
\item The black hole starts to evaporate in a discrete way; for simplicity we assume that each emitted quantum of the field has the same energy $\eps$, so that $M_0 = N_G \,\eps$ for some integer $N_G$.
\item At the end of the evaporation, the space becomes almost flat again and the field is in excited state with $N_G$ quanta.
\end{enumerate}
We assume:
\begin{enumerate}
\item There exists a \emph{fundamental Hilbert space} $\HH$. That is the Hilbert space of the fundamental degrees of freedom of the total system,
  i.e.,\ black hole, radiation and space outside the black hole. Since here we focus on a finite region accessible to a generic observer and big enough to contain the
  black hole at initial time and the emitted radiation at a later time, $\HH$ here is finite-dimensional;

\item For a specific observer at low-energy scale, the states of $\HH$ appear as classical \emph{spatial} geometry and quantum fields propagating on it.

\item There are states in $\HH$ which represent the same classical geometry but are microscopically different.

\item In general, there is exchange of the number of degrees of freedom between the fields and geometry.
\end{enumerate}

In this model, we introduce a space of classical geometries representing spatial slices of space-time containing a black hole of a given mass $M^{(a)} = a\,\eps$. That is, we introduce an orthonormal set of
the states
\begin{align}\label{eq:ga state}
  \ket{ g^{(a)} }, \qquad a = 0, 1, \dots N_G - 1, 
\end{align}
where $N_G$ is therefore the number of geometries allowed in our model. For convenience, we introduce the \emph{Hilbert space of classical geometries} $\HHG$ as the linear span of the states (\ref{eq:ga state}) and
define the ``mass operator'' ${\MM}$ by
\begin{align}
  {\MM} \ket{g^{(a)}} =  M^{(a)} \ket{g^{(a)}} \equiv \eps\,a\,\ket{g^{(a)}}.
\end{align}
An operator of this kind should represent the possibility of measuring geometric properties of the space, such as the three-dimensional metric, as seen
by a specific observer.  The assumption that the geometry of the space is a result of some coarse-graining procedure associated with a specific observer
means there is some mapping $\PPG: \HH \mapsto \HHG$ which assigns to a microscopic state in $\HH$ corresponding classical geometry
or an appropriate superposition of such geometries.

Similarly, we shall assume the existence of some mapping $\PPF:\HH \mapsto \HHF$ which extracts the ``field content'' of a state in $\HH$. Then, $\HHF$ can be, e.g.,\ an appropriate
Hilbert (Fock) space representing the states of the fields; concrete definitions will depend on the
particular theory of quantum gravity. Schematically, the states of the fundamental Hilbert space $\HH$ can be interpreted as states with some classical geometry
via the mapping $\PPG$, and with some state of the quantum field via the mapping $\PPF$:
\begin{center}
\begin{tikzpicture}
  \draw (0,0) node {$\ket{\psi}\in\HH$};
  \draw[->] (-0.5,-0.3)--+(-1.5,-0.5) node[pos=0.75,above=3pt] {\scriptsize $\PPG$};
  \draw[->] (0.3,-0.3)--+(1.5,-0.5) node[pos=0.75,above=3pt] {\scriptsize $\PPF$};
  \draw (-2, -1.1) node { $\ket{g^{(a)}}\in\HHG$};
  \draw (2, -1.1) node { $\ket{\phi} \in \HHF$};  
\end{tikzpicture}
\end{center}
After introducing these mappings, one can label the states in $\HH$ by the values of the coarse-grained quantities, i.e., $\ket{\psi}=\ket{g^{(a)}, \phi}$.

For simplicity we assume that any state of $\HH$ can be interpreted in such a way, although in reality
this is much more complicated: classical geometries are expected to be very special superpositions of basis
states with no classical analogues. Since we are not building
a specific model of quantum gravity, we ignore this complication. On the other hand, one can argue that among the states corresponding to
definite classical geometries one can choose a subset
of (sufficiently distinct) states which are approximately orthogonal and consider
only a subspace of $\HH$ generated by this (approximately) orthonormal set.

In Page's picture described in the Sec.\ \ref{sec:page-curve} he considers
splitting of the Hilbert space representing the states of the field into ``inside'' and ``outside'' part with respect
to the horizon of the black hole. In our model we wish to implement the idea that the
geometry and its fundamental degrees of freedom must be brought into the picture, so that one should
split the fundamental space $\HH$ into a direct product of ``geometrical'' and ``field'' part. However, for our argument it is essential
to entertain the possibility that the distribution
of the microscopic degrees of freedom between the geometry and the fields is not fixed and can change during the evolution of the system.

The following simple model will serve just as a useful visualization and to provide a
terminology convenient for the subsequent construction. However, the construction itself does not rely on such
visualization. Let there be a certain number $N$ of fundamental degrees of freedom in the sense explained in the Introduction.  The states
of each fundamental degree of freedom form a $d$-dimensional Hilbert space, so that the Hilbert space of all fundamental degrees of freedom has dimension $d^N$. Now,
the states of the fundamental degrees of freedom give rise to
the notions of spatial geometry (distance, topology, dimension) and of quantum fields. We think of
the set of all fundamental degrees of freedom as distributed among the vertices of a graph and their links.  More specifically, suppose
that the fact that there is some geometrical relation (e.g.,\ the distance) between two vertices can
be represented as a link between corresponding vertices of the graph and a quantitative
measure of such relation is represented by a weight (or a set of weights) of the link. Thus, one could interpret the geometry as encoded
in the states of all links in the graph. However, in order to keep the total degrees of
freedom constant, the vertices have to ``offer'' some of their degrees of freedom to form the Hilbert space $\HHG$
corresponding to the states of the links which are interpreted as the state of the geometry. Then, the
remaining fundamental degrees of freedom can be represented as excitation states of the vertices
of the graph and they form a Hilbert space $\HH_F$ whose elements are interpreted as the
states of the emergent field. So, the state of the entire graph is an element of the Hilbert space $\HHG \otimes \HHF$
of dimension $d^N$. The point is that it is the \emph{topology} of the graph (by which we mean simply a specific distribution
of the links, ignoring their weights and the states of the vertices) which
dictates how the available fundamental degrees of freedom are distributed between the fields and the
geometry. During a standard, ``nonviolent'' evolution,
we might expect that the topology of the lattice
does not change, but as the black hole and singularity form, significant changes
of the topology happen, implying both topological and causal changes in the emergent
spatial geometry and possible deviations from standard QFT on curved space-time in the following
sense: the change of the topology of the lattice means reshuffling of the fundamental degrees of
freedom between the geometry and the fields,  so that the structure of the new graph
is $\HHG' \otimes \HHF'$; the fields now live in a Hilbert space $\HHF'$ of different dimension
than $\HHF$. In this case we have to expect the deviations from the unitary evolution on the effective field side, although
the underlying evolution of microscopic degrees of freedom is purely unitary.

We do not stick to this oversimplified picture in which the weights of the links are related directly to the
metric and the states of the vertices are related directly to the states of the fields. We
shall, however, stick to the idea that there are several ways how the fundamental degrees  of
freedom are reshuffled between the fields and the geometry and, in addition, there might exist
different microscopic configurations which, on the effective level, give rise to the same coarse-grained geometry. On
the effective level it is impossible to distinguish between two such
microscopic configurations but, microscopically, the two configurations differ by the number of degrees
of freedom available for the fields. That is, the fields in the two cases are
elements of different Hilbert spaces and, hence, the resulting field cannot be in a pure state.

\subsection{Toy model}

Hence, starting from the fundamental Hilbert space $\HH$, we assume it can be split into a direct sum of the subspaces $T_{(i)}$,
\begin{align}
  \HH &= \bigoplus_{i=1}^{N_T} T_{(i)}, \qquad \dim \HH = N_T \, N,
\end{align}
where each $T_{(i)}$ has a fixed dimension $N$ and consists of states with some specific distribution of the degrees of freedom between the geometry and the
fields; in the language of the simplistic ``graph model'', $T_{(i)}$ is a set of states for one specific choice of the topology of the graph and hence
we shall refer to $T_{(i)}$ as the set of the states with specific topology; $N_T$ is then the number of different topologies. By assumption, each $T_{(i)}$ has a structure
\begin{align}
  T_{(i)} &= \HH_{\mathrm{G}}^{p_i} \otimes \HH_{\mathrm{F}}^{q_i}, \qquad p_i\,q_i = N,
\end{align}
where $\HH_{\mathrm{G}}^{p}$ ($\HH_{\mathrm{F}}^q$) is a Hilbert space of dimension $p$ ($q$) representing possible microscopic states of the geometry (fields).

A general state $\ket{\psi}\in\HH$ admits the expansion adapted to the splitting of $\HH$ which is in the form
\begin{align}
  \ket{\psi} &= \bigoplus_{i=1}^{N_T} \sum_{I=1}^{p_i} \sum_{n=0}^{q_i-1} c^{(i)}_{In} \ket{I_i} \otimes \ket{n_i},
\end{align}
where vectors $\ket{I_i}$ and $\ket{n_i}$ form a basis of spaces $\HH_{\mathrm{G}}^{p_i}$ and $\HH_{\mathrm{F}}^{q_i}$, respectively. 

Let us denote by ${\PP}_{(i)} : \HH \mapsto T_{(i)}$ a projector onto the subspace $T_{(i)}$. Then, the squared norm of the state ${\PP}_{(i)}\ket{\psi}$
is the probability $p_{(i)}$ of finding the system in the state with the topology $T_{(i)}$,
\begin{align}
  p_{(i)} &= \| {\PP}_{(i)}\ket{\psi}\|^2.
\end{align}
In general, state in $T_{(i)}$ is a state with the entanglement between the geometry and the field in the sense that its decomposition reads
\begin{align}
  {\PP}_{(i)}\ket{\psi} &= \sum_{I,n} c^{(i)}_{In} \ket{I_i} \otimes \ket{n_i}.
\end{align}
Associated density matrix representing the state of the field is
\begin{align}\label{eq:rho-i}
  {\rho}_{(i)} &= \Tr_{\HH_{\mathrm{G}}^{p_i}} \ket{\psi}_{i} \bra{\psi}_i ,
\end{align}
where we first define the normalized state
\begin{align}
  \ket{\psi}_i &= p_{(i)}^{-1/2} {\PP}_{(i)}\ket{\psi}
\end{align}
and then trace over the degrees of freedom of the gravitational field. Corresponding entanglement entropy will be denoted by
\begin{align}\label{eq:S-i}
  S_{(i)} &= - \Tr_{\HH_{\mathrm{F}}^{q_i}} {\rho}_{(i)} \ln {\rho}_{(i)};
\end{align}
$S_{(i)}$ is the entanglement entropy between the geometry and the fields for a given topology of the lattice. Since for the observer it is impossible to distinguish between different topologies
of the lattice, expected value of the entanglement between the fields and the geometrical degrees of freedom will be
\begin{align}\label{eq:S avg}
  \langle S \rangle &= \sum_i p_{(i)} S_{(i)}.
\end{align}

In our toy model we shall assume that only two topologies are possible, i.e.,\ $N_T = 2$, and that both topologies admit the same family of classical geometries (\ref{eq:ga state}), i.e.,\ we assume
\begin{align}
  \PPG( T_{(1)} ) &= \PPG( T_{(2)}) = \HHG.
\end{align}
Let us fix the number of degrees of freedom for each type of the lattice to $N = 1500$ and let us set
\begin{align}
  T_{(1)} &= \HH_{\mathrm{G}}^{30} \otimes \HH_{\mathrm{F}}^{50}, &
                                              p_1 \times q_1 &= 30 \times 50, \nonumber \\
  T_{(2)} &= \HH_{\mathrm{G}}^{60} \otimes \HH_{\mathrm{F}}^{25}, &
                                             p_2 \times q_2 &= 60 \times 25;
\end{align}
then we have $\dim \HH = 3000$. Finally, assume that the maximal mass $M_0$ of the black hole is split into $N_G = 30$ quanta. That is, for a black
hole of mass $M^{(a)}=a\,\eps$ there is exactly one state in $\HH_{\mathrm{G}}^{30}$ such that the mapping $\PPG$ maps it to a state $\ket{g^{(a)}}$,
while in $\HH_{\mathrm{G}}^{60}$ there are two such states. Hence, the subspaces $T_{(i)}$ are generated by the following bases:
\begin{align}
  T_{(1)} &= \Span\left\{ \ket{{(a)}} \otimes \ket{n} \right\}, \nonumber \\
  T_{(2)} &= \Span\left\{ \ket{(a),1}\otimes \ket{n}, \ket{(a),2}\otimes \ket{n} \right\},
\end{align}
where
\begin{align}
  \PPG &: \ket{(a)}\otimes \ket{n} \in \HH_{\mathrm{G}}^{(30)} \otimes \HH_{\mathrm{F}}^{(50)}& \mapsto& \ket{g^{(a)}} \in \HHG, \nonumber \\
 &: \ket{(a),1} \otimes \ket{n} \in \HH_{\mathrm{G}}^{(60)} \otimes \HH_{\mathrm{F}}^{(25)}& \mapsto & \ket{g^{(a)}} \in \HHG, \nonumber \\
 &:   \ket{(a),2} \otimes \ket{n} \in \HH_{\mathrm{G}}^{(60)} \otimes \HH_{\mathrm{F}}^{(25)} & \mapsto &  \ket{g^{(a)}} \in \HHG.
\end{align}
We shall interpret elements  $\ket{n} \in \HH^q_{\mathsf{F}}$ as the states of the quantum field with $n$ particles.

A general state in $\HH$ can be now written in the form
\begin{multline}\label{eq:general psi}
  \ket{\psi} = \doublesumone \alpha_{an} \ket{(a)}\otimes \ket{n} + \\
             + \doublesumtwo \left( \beta_{an} \ket{(a),1}\otimes \ket{n} + \gamma_{an} \ket{(a),2}\otimes \ket{n} \right),
\end{multline}
and the expected value of the mass of the black hole is
\begin{multline}\label{eq:average}
  \avg{M} = \sum_{a=0}^{N_G-1}\sum_{n=0}^{q_1-1} M^{(a)} |\alpha_{an}|^2 +\\
          +  \sum_{a=0}^{N_G-1}\sum_{n=0}^{q_2-1} M^{(a)} (|\beta_{an}|^2 + |\gamma_{an}|^2),
\end{multline}
and similarly for the expected number of particles $\avg{n}$. The probabilities for finding the lattice in the state with the topology $T_{(1)}$ and $T_{(2)}$, respectively, are
\begin{align}
  p_{(1)} &=\doublesumone |\alpha_{an}|^2, \\
  p_{(2)} &=\doublesumtwo \left( |\beta_{an}|^2 + |\gamma_{an}|^2 \right).
\end{align}

During the evolution (first formation of the black hole, then evaporation) the system evolves continuously and unitarily in $\HH$. We
start the analysis at the point when the black hole just formed and the field outside the black hole is in the ground state, i.e.,\ in
the state with zero particles. Then the evaporation starts which we mimic by prescribing the expected values of mass $\avg{M}$ and the
expected value of number of particles $\avg{n}$. Since we do not know the underlying microscopic dynamics, we shall consider, similarly
to Page, all states which are compatible with these expectation values.

Let us find a convenient parametrization of such states. First, since the general state (\ref{eq:general psi}) must be normalized,
\begin{align}
\doublesumone |\alpha_{an}|^2 + \doublesumtwo (|\beta_{an}|^2 + |\gamma_{an}|^2) = 1
\end{align}
we can set
\begin{align}
p_{(1)} &=  \sum_{n=0}^{q_1-1} |\alpha_{an}|^2 = \cos^2\theta, \nonumber\\
p_{(2)} &=  \sum_{n=0}^{q_2-1}( |\beta_{an}|^2 + |\gamma_{an}|^2) = \sin^2\theta,
\end{align}
where $\theta \in (0, \pi/2)$ without the loss of generality. Let us parametrize the coefficients by
\begin{align}\label{eq:modul alpha beta gamma}
  \begin{split}
    |\alpha_{an}| &= \mu_{an}  \cos\theta, \\
    |\beta_{an}| &= \nu_{an} \sin\theta \cos\phi, \\
    |\gamma_{an}| &= \lambda_{an} \sin\theta \sin\phi,
  \end{split}
\end{align}
where $\phi \in (0,\pi/2)$. This is analogous to introducing the Hopf coordinates on the sphere $S^n$. This parametrization implies
\begin{align}
  \doublesumone \mu_{an}^2 = \doublesumtwo\nu_n^2 =\doublesumtwo \lambda_n^2 = 1.
\end{align}
We also define
\begin{align}\label{eq:mu-bar}
  \mu_M &= \doublesumone M^{(a)}\,\mu_{an}^2, &
                                             \mu_N &= \doublesumone n\,\mu_{an}^2
\end{align}
and similarly for $\nu_M, \nu_N$ and $\lambda_M, \lambda_N$. In this notation, the expected value of the mass of the black hole and the expected number of particles are given by
\begin{align}
  \begin{split}
  \avg{M} &= \mu_M\cos^2\theta + \sin^2\theta(\nu_M \cos^2\phi + \lambda_M \sin^2\phi),\\
  \avg{n} &= \mu_N\cos^2\theta + \sin^2\theta(\nu_N \cos^2\phi + \lambda_N \sin^2\phi).
\end{split}\label{eq:avg-n}
\end{align}
A generic state (\ref{eq:general psi}) now acquires the form
\begin{multline}
    \ket{\psi} = \cos\theta \doublesumone \mu_{an} e^{\ii \chi^\alpha_{an}} \ket{(a)}\otimes\ket{n} +\\
    +\sin\theta  \doublesumtwo \left( \nu_{an} e^{\ii \chi^\beta_{an}}\cos\phi \ket{(a),1}\otimes\ket{n} + \right.\\
\left. + \lambda_{an} e^{\ii \chi^\gamma_{an}} \sin\phi \ket{(a),2}\otimes\ket{n} \right).\label{eq:psi}
\end{multline}
The normalized projections of $\ket{\psi}$ onto $T_{(1)}$ and $T_{(2)}$ correspond to the first and the second sum in (\ref{eq:psi}), respectively, with factors $\sin\theta$ and $\cos\theta$ omitted:
\begin{subequations}
\begin{align}
  \ket{\psi}_1 &= \doublesumone \mu_{an}\,e^{\ii\chi^\alpha_{an}} \ket{(a)} \otimes \ket{n}, \\
  \ket{\psi}_2 &= \cos\phi \doublesumtwo \nu_{an}\,e^{\ii\chi^\beta_{an}} \ket{(a),1} \otimes \ket{n} + \nonumber \\
  & + \sin\phi \doublesumtwo \lambda_{an}\,e^{\ii\chi^\gamma_{an}} \ket{(a),2} \otimes \ket{n}.
\end{align}\label{eq:psi-1-2}
\end{subequations}
Corresponding density matrices are
\begin{align}
  \rho_{(i)} &= \sum_{m,n=0}^{q_i-1} c^{(i)}_{nm} \,\ket{n} \bra{m}, \quad i=1,2,
\end{align}
where
\begin{align}
  \begin{split}
    c^{(1)}_{mn} &= \sum_{a=0}^{N_G-1} \mu_{an}\,\mu_{am}\,e^{\ii \chi^\alpha_{an}-\ii \chi^\alpha_{am}}, \\
    c^{(2)}_{mn} &= \sum_{a=0}^{N_G-1} \left( \nu_{an}\,\nu_{am}\,e^{\ii \chi^\beta_{an}-\ii \chi^\beta_{am}}
      +  \lambda_{an}\,\lambda_{am}\,e^{\ii \chi^\gamma_{an}-\ii \chi^\gamma_{am}}\right).
  \end{split}\label{eq:c-1-2}
\end{align}
With each density matrix $\rho_{(i)}$ there is associated entanglement entropy $S_{(i)}$ given by (\ref{eq:S-i}), so that the average value of the entanglement entropy is (\ref{eq:S avg})
\begin{align}\label{eq:S avg theta}
  \avg{S} &= S_{(1)}\cos^2 \theta +S_{(2)} \sin^2\theta.
\end{align}
For the purposes of this paper, the corresponding entanglement entropy will be calculated numerically. 

\subsection{Entanglement entropy calculation}

We wish to estimate the expected entanglement entropy for a random state which has specific expected value
of number of particles $\avg{n}$. In order to do that, we would need to solve the constraints (\ref{eq:avg-n}) for given values $\avg{M}$ and $\avg{n}$ with respect to
the parameters $\theta$, $\phi$, $\mu_M$, $\mu_N$, $\nu_M$, $\nu_N$ and $\lambda_M$, $\lambda_N$. For the purposes of this paper we chose the following way of estimating the entanglement
entropy.

Conditions (\ref{eq:avg-n}) can be rewritten in the form
\begin{align}
  x^2 &= \frac{\mu_M}{\avg{M}} \cos^2 \theta, & x'^2 &= \frac{\mu_N}{\avg{n}} \cos^2 \theta, & \nonumber\\
  y^2 &= \frac{\nu_M}{\avg{M}} \sin^2\theta \cos^2 \phi, & y'^2 &= \frac{\nu_N}{\avg{n}} \sin^2\theta \cos^2 \phi, & \\
  z^2 &= \frac{\lambda_M}{\avg{M}} \sin^2\theta \sin^2\phi,&
  z'^2 &= \frac{\lambda_N}{\avg{n}} \sin^2\theta \sin^2\phi,\nonumber
\label{eq:xyz}
\end{align}
where $(x,y,z)$ and $(x',y',z')$ are points on the unit 2-sphere. Hence, we generate two random unit 3-vectors and choose $\cos^2\theta$ and $\cos^2\phi$ randomly  with the uniform
distribution on the interval $(0, 1)$ and check, whether conditions
\begin{align}
 \frac{\avg{M}\,x^2}{\cos^2\theta} & \leq N_G-1, &  \frac{\avg{n}\,x'^2}{\cos^2\theta} & \leq q_1-1, & \\
  \frac{\avg{M}\,y^2}{\sin^2\theta \cos^2\phi} & \leq N_G-1, &   \frac{\avg{n}\,y'^2}{\sin^2\theta \cos^2\phi} & \leq q_2-1, & \\
  \frac{\avg{M}\,z^2}{\sin^2\theta \sin^2\phi} & \leq N_G-1, & \frac{\avg{n}\,z'^2}{\sin^2\theta \sin^2\phi} & \leq q_2-1,
\end{align}
are satisfied, where, we recall, $N_G$ is the number of allowed geometries and $q_1 = \dim \HH_{\mathrm{F}}^{q_1}=30, q_2 = \dim \HH_{\mathrm{F}}^{q_2}=60$ are the dimensions of
Hilbert spaces for the fields corresponding to the two topologies of the lattice. If the conditions
do not hold, we iterate the procedure until we find such combination of $\theta, \phi$, $(x,y,z)$ and $(x',y',z')$. This ensures that
sequences $\mu_{an}$, $\nu_{an}$ and $\lambda_{an}$ with the desired averages $\mu_M, \mu_N, \nu_M, \nu_N, \lambda_M$ and $\lambda_N$ exist. Then we generate
such sequences. Finally, we choose the phases $\chi^\alpha_{an}$, $\chi^\beta_{an}$ and $\chi^\gamma_{an}$ randomly with uniform distribution on the interval $(0,2\pi)$.
In this way we generate a random state yielding the prescribed expectation values $\avg{M}$ and $\avg{n}$; the entropy is then calculated by means of Eqs.\ \eqref{eq:S-i} and
\eqref{eq:c-1-2}. Then we calculate the average value of entanglement entropy from 5000 runs of this procedure. 

Now we consider the scenario of evaporation of the black hole. At the beginning, let the black hole have its maximal mass $M_0 = (N_G -1) \eps$, where we choose $N_G=30$ and let
there be vacuum outside the black hole. That does not necessarily mean that the Hilbert space for the field outside
the black hole has dimension 1 (as it is in Page's case); indeed, in our model we have chosen the
dimension to be either 50 or 25, depending on the topology of the lattice. However, since there is only one vacuum
state $\ket{0} \in \HH_{\mathrm{F}}^{50}$ and only one vacuum state $\ket{0}\in \HH_{\mathrm{F}}^{25}$,
in both topologies, the field is disentangled from the geometry. This can be also seen from Eq.\ (\ref{eq:avg-n}) which shows that for $\avg{n}=0$ we have
\begin{align}
\mu_N = \nu_N = \lambda_N = 0
\end{align}
which, by (\ref{eq:mu-bar}), implies
\begin{align}
  \mu_{an} &=0 \quad \text{for}~n > 0,
\end{align}
and similarly for $\nu_{an}$ and $\lambda_{an}$. Requirement $\mu_M = M_0$ then implies that the
only nonzero $\mu_{an}$ is
\begin{align}
  \mu_{(N_G-1),0} &= 1.
\end{align}
Then both states $\ket{\psi}_1$ and $\ket{\psi}_2$ in (\ref{eq:psi-1-2}) are unentangled and we have $\avg{S}=0$. Hence, our starting point coincides with the starting point of Page: expected entanglement vanishes at the beginning of the evaporation.

Now we assume that black hole starts to evaporate. We assume continuous unitary evolution of the state
in $\HH$ but take the ``snapshots'' of the system when the expected values are
\begin{align}
  \avg{M} &= (N_G-1-k)\,\eps, &
                        \avg{n} &= k,
\end{align}
where $k$ acquires discrete values
\begin{align}
  k &= 0, 1, \dots N_G-1;
\end{align}
$k=0$ corresponds to black hole of maximal mass $M_0=M^{(N_G-1)}=(N_G-1)\eps$ and vacuum outside
the black hole; in $k$-th step, black hole already emitted $k$ quanta of the field, so its mass
decreased by the value $k\,\eps$, while the field is in the state with $k$ quanta outside; black
hole is fully evaporated for $k=N_G-1$ and the field is in the state with $N_G-1$ particles.

Notice that at the end of the evaporation, the state $\ket{\psi}_1$ is disentangled again, but the
state $\ket{\psi}_2$ remains entangled. Hence, the expected value of the entanglement entropy decreases
but remains nonzero.

\begin{figure}
  \centering
  \includegraphics[width=0.45\textwidth]{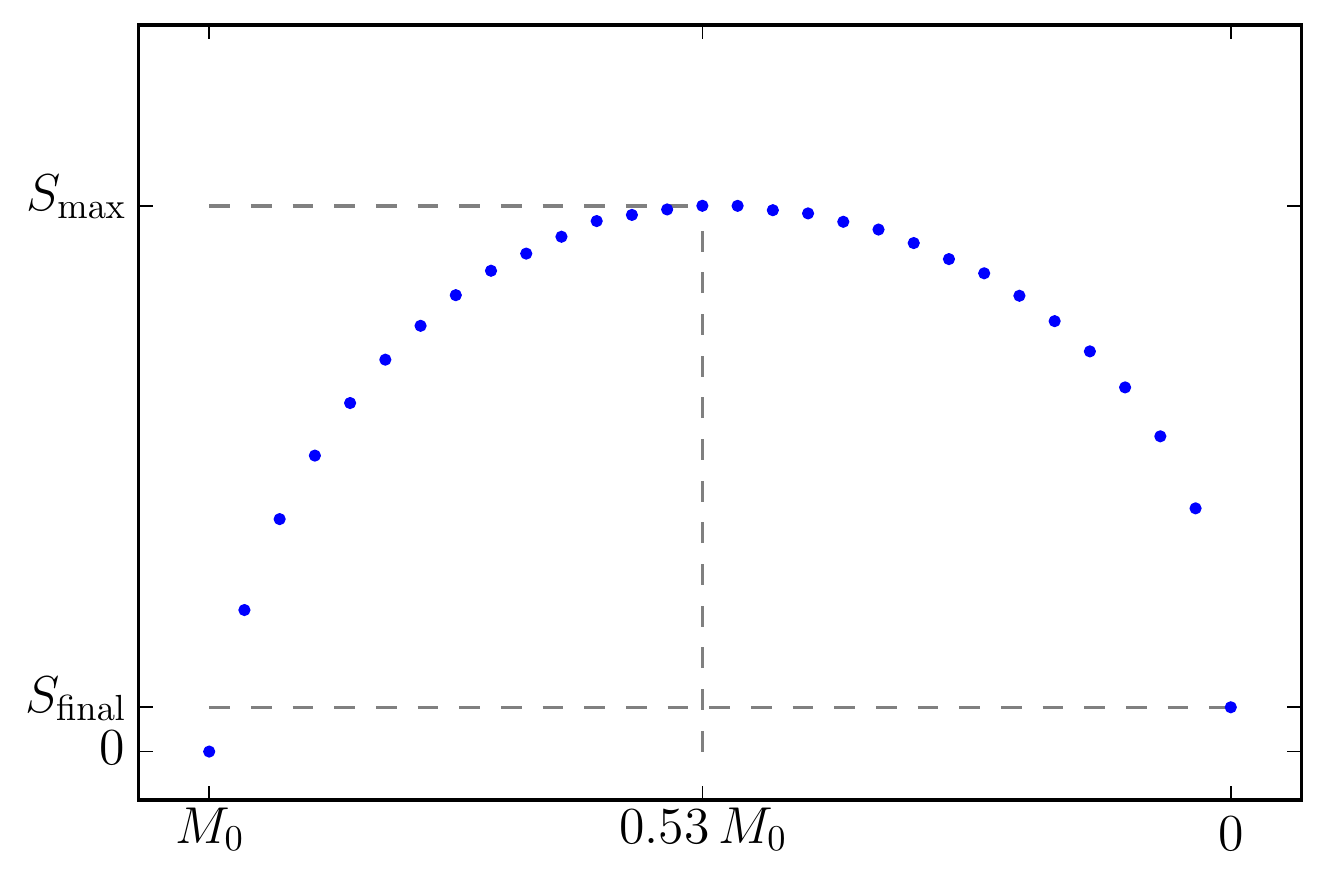} 
  \caption{Entanglement entropy as a function of the decreasing mass of the black hole during the process of evaporation.  Values of the parameters are given in the main text.}
  \label{fig:page-modified}
\end{figure}

In Fig.\ \ref{fig:page-modified} we show the entanglement entropy as the function of the discrete
parameter $k$. Although this graph starts at the point $(M_0,0)$ which corresponds to the same origin of the Page curve in Fig.\ \ref{fig:page-original}, at the final stage of the evaporation the entanglement entropy does not go to zero; at this point we differ from the prediction of the Page curve. It is clear that allowing for more microscopic realizations of the same effective geometry, i.e.,\
more topologies, would in general increase the final deviation of $\avg{S}$ from the pure state value.

\bigskip

\section{Conclusions}\label{sec:conc}

The fact that the number of degrees of freedom that determine the state of a system in a compact volume $V$ is bounded from above by the Bekenstein-Hawking entropy of a black hole with horizon area $\partial V$, leads us to entertain the possibility that such fundamental degrees of freedom should describe the state of both fields {\it and} geometry contained in said volume.  The immediate consequence of such statement is that fields and geometry should be regarded as emergent phenomena at ordinary energy scales.  In this picture the particles of the Standard Model are regarded as analogous to quasiparticles, arising together with the classical background geometry from the interactions between the fundamental degrees of freedom.  We have not provided here a framework for such dynamical emergence (see \cite{Cao2016,maldacena2013,jensen2013,Konopka2006,Baez1999,Ambjorn2006,Lombard2016,Oriti2014,Rastgoo2016,Requardt2015,Trugenberger2016} for some examples along these lines); however we assume that the unitary evolution -- which is a central requirement of the emergent quantum theory -- is preserved down to the fundamental level.  The unitary evolution inevitably leads to a {\it reshuffling} of the fundamental degrees of freedom and this is reflected on the emergent level as an entanglement between quantum fields and geometry.  In order to investigate the consequences of such scenario, we provided here a kinematical framework that allows us to address the longstanding information-loss paradox in the context of black hole evaporation.  Through a simple toy model of evaporation it is shown how the entanglement between fields and geometry can lead, after the evaporation is completed, to an average loss of the initial infomation.  We claim that such modification of the original Page curve should be regarded as a common feature of any theory of quantum gravity in which both the spacetime geometry and the quantum fields propagating on it are emergent features of an underlying fundamental and unitary theory.

\section*{Acknowledgments}

A.I.\ acknowledges partial financial support from the Czech Science Foundation (GA\v{C}R), under contract no.\ 14-07983S. The work of G.A.\ and M.S.\ is financially supported by the Czech
Science Foundation, grant GA\v{C}R no.\ 17-16260Y. All authors are indebted to Georgios Luke\v{s}-Gerakopoulos and Pavel Krtou\v{s} for their critical remarks and many fruitful and inspirational discussions.

%

\end{document}